\documentclass[graybox]{svmult}

\nonstopmode
\usepackage{mathptmx}       
\usepackage{helvet}         
\usepackage{courier}        
\usepackage{makeidx}         
\usepackage{graphicx}        
\usepackage[bottom]{footmisc}
\usepackage{url}
\usepackage[utf8]{inputenc}
\usepackage[numbers,sort]{natbib}
\usepackage{hyperref}
\usepackage{stmaryrd}
\usepackage{amssymb}
\usepackage{amsmath}
\usepackage{listings}

\usepackage{tikz}

\usetikzlibrary{positioning}
\usetikzlibrary{calc}
\usetikzlibrary{chains}
\usetikzlibrary{shapes.misc}
\usetikzlibrary{shapes.symbols}
\usetikzlibrary{shapes.geometric}
\usetikzlibrary{fit}
\usetikzlibrary{shadows}
\usetikzlibrary{arrows}
\usetikzlibrary{trees}
\usetikzlibrary{decorations.pathmorphing}

\pgfdeclarelayer{background}
\pgfdeclarelayer{foreground}
\pgfsetlayers{background,main,foreground}

\newcommand{\doi}[1]{\href{http://dx.doi.org/#1}{doi: #1}}
\newcommand{\mathd}{\,\mathrm{d}}
\renewcommand{\D}{\mathsf{D}}
\newcommand{\assign}{:=}
\colorlet{codeback}{gray!20}

\newcommand{\jump}[1]{\left\llbracket#1\right\rrbracket}

\begin{document}

\title{High-Order Discontinuous Galerkin Methods by GPU
Metaprogramming}


\author{Andreas Klöckner, Timothy Warburton and Jan S. Hesthaven }
\institute{%
Andreas Klöckner \at Courant Institute of Mathematical Sciences, New York University, New York, NY 10012, \email{kloeckner@cims.nyu.edu}
\and Tim Warburton \at Department of Computational and Applied Mathematics, Rice University, Houston, TX 77005 \email{timwar@caam.rice.edu}
\and Jan S. Hesthaven \at Division of Applied Mathematics, Brown University, Providence, RI 02912 \email{jan.hesthaven@brown.edu}}

\maketitle


\abstract{
Discontinuous Galerkin (DG) methods for the numerical solution of
partial differential equations have enjoyed considerable success
because they are both flexible and robust: They allow arbitrary
unstructured geometries and easy control of accuracy without
compromising simulation stability. In a recent publication, we have
shown that DG methods also adapt readily to execution on modern,
massively parallel graphics processors (GPUs).  A number of qualities
of the method contribute to this suitability, reaching from locality
of reference, through regularity of access patterns, to high
arithmetic intensity. In this article, we illuminate a few of the more
practical aspects of bringing DG onto a GPU, including the use
of a Python-based metaprogramming infrastructure that was created
specifically to support DG, but has found many uses across all
disciplines of computational science.
}

\section{Introduction}
\label{sec:intro}

Discontinuous Galerkin methods
\cite{reed_triangular_1973,lesaint_finite_1974,cockburn_runge-kutta_1990,hesthaven_nodal_2007}
are, at first glance, a rather curious combination of ideas from
Finite-Volume and Spectral Element methods. Up close, they are very much
high-order methods by design.  But instead of perpetuating the order
increase like conventional global methods, at a certain level of detail,
they switch over to a decomposition into computational elements and
couple these elements using Finite-Volume-like surface Riemann solvers.
This hybrid, dual-layer design allows DG to combine advantages from both
of its ancestors.  But it adds a third advantage: By adding a movable
boundary between its two halves, it gives implementers an added degree
of flexibility when bringing it onto computing hardware.

Using graphics processors for computational tasks is by no means a new
idea.  In fact, even in the days of marginally programmable
fixed-function hardware, some (especially particle-based) methods
obtained large performance gains from running on early GPUs.  (e.g.
\cite{li_implementing_2003}) In the domain of solvers for partial
differential equations, Finite-Difference Time-Domain (FDTD) methods
are a natural fit to graphics processors and obtained high performance
with relative ease (e.g., \cite{krakiwsky_acceleration_2004}). Finite
Element solvers were also brought onto GPUs relatively early on (e.g.,
\cite{goeddeke_accelerating_2005}), but often failed to reach the same
impressive speed gains observed for the simpler FD methods. In the
last few years, high-level abstractions such as Brook and Brook for
GPUs \cite{buck_brook_2004} have enabled more and more complex
computations on streaming hardware. Building on this work, Barth et
al.\ \cite{barth_streaming_2005} already predicted promising
performance for two-dimensional DG on a simulation of the Stanford
Merrimac streaming architecture \cite{dally_merrimac_2003}.  Nowadays,
compute abstractions are becoming less encumbered by their graphics
heritage \cite{lindholm_nvidia_2008, cuda_prog_2008}. This has helped
bring algorithms of ever higher complexity onto the GPU. Taking
advantage of these advances, our paper \citep{kloeckner_nodal_2009}
presented, to the best of our knowledge, the first implementation of a
discontinuous Galerkin method on a single real-world consumer graphics
processor. Now, a few years after the publication of the original
paper, interest in GPUs and their use for solving partial differential
equations continues unabated. A few implementers have followed in our
footsteps and brought their versions of DG onto GPUs.

Let us briefly place this text within the sequence of articles on
GPU-DG we have authored. The first one \citep{kloeckner_nodal_2009} is
rather technical and introduces all the tricks and details needed to
make the method go fast. The second one \citep{kloeckner_solving_2011}
serves as an introduction to be read by a larger, somewhat
non-technical audience. This latest one addresses some of the software
challenges involved in achieving fast execution of GPU-based
discontinuous Galerkin methods.

The article is structured as follows: In Section
\ref{sec:dg-overview}, we review the details of the discontinuous
Galerkin method and its implementation in general, followed by a
discussion of considerations required by its implementation on GPUs
specifically in Section \ref{sec:gpu-dg}. Responding to the challenges
of this section, we introduce the motivation and implementation details of
our Python-based infrastructure for run-time code generation (RTCG) in
Section \ref{sec:rtcg}. We then discuss the specifics of RTCG in the
context of DG in Section and confirm the success of the method through
experimental results in Section \ref{sec:results}. In closing and
summing up what was achieved, we outline avenues for future work in
Section \ref{sec:conclusions}.

\section{The Discontinuous Galerkin Method}

\label{sec:dg-overview}

By their design and origins, DG methods are particularly suited to
approximating the solution of a hyperbolic system of conservation laws
\begin{equation} 
  u_t + \nabla \cdot F (u) = 0. \label{eq:claw} 
\end{equation}
Initial boundary value problems for PDEs that can be cast in the form
\eqref{eq:claw} as well as slight generalizations thereof, include
Maxwell's equations, Euler's equations of gas dynamics, the
Navier-Stokes equations, equations arising from Lattice-Boltzmann
models, the equations of magnetohydrodynamics, or the shallow-water
equations.  In summary, a wide variety of physical phenomena in the
time domain can be modeled using this type of equation.

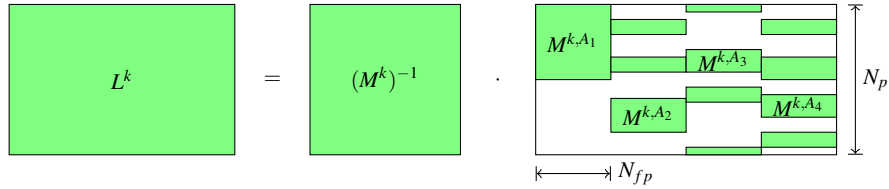
\begin{figure}
\begin{center}
\begin{tikzpicture}[
  densemat/.style={fill=green!50},
  ]
  \draw [densemat] (-1,-1) rectangle (1,1) ;
  \draw (2,-1) rectangle +(4,2) ;
  \draw [densemat] (-2,-1) rectangle (-5,1) ;

  \draw [densemat] (2,1) rectangle +(1,-1) ;
  \foreach \ys/\ye in {0.1/0.3, -0.25/-0.7, 0.6/0.8}
  { \draw [densemat] (3,\ys) rectangle (4,\ye) ; }
  \foreach \ys/\ye in {0.4/0.1, -0.1/-0.3, -0.9/-1, 1/0.9}
  { \draw [densemat] (4,\ys) rectangle (5,\ye) ; }
  \foreach \ys/\ye in {0.8/0.6, 0.3/0, -0.9/-0.7, -0.2/-0.5}
  { \draw [densemat] (5,\ys) rectangle (6,\ye) ; }

  \node at (2.5,0.5) {$M^{k,A_1}$} ;
  \node at (3.5,-0.475) {$M^{k,A_2}$} ;
  \node at (4.5,0.25) {$M^{k,A_3}$} ;
  \node at (5.5,-0.35) {$M^{k,A_4}$} ;

  \node at (0,0) {$(M^k)^{-1}$} ;
  \node at (1.5,0) {$\cdot$} ;
  \node at (-1.5,0) {$=$} ;
  \node at (-3.5,0) {$L^k$} ;

  \draw [|<->|] (6.25,-1) -- +(0,2) node [pos=0.5, anchor=west] {$N_p$} ;

  \draw [|<->|] (2,-1.25) -- (3,-1.25) node [anchor=west] {$N_{fp}$} ;
\end{tikzpicture}
\end{center}
\caption{Construction of the Lifting Matrix $L^k$.}
\label{fig:lifting-matrix}
\end{figure}

\eqref{eq:claw} is to be solved on a domain $\Omega =
\biguplus_{k=1}^K \D_k \subset \mathbb R^d$ consisting of disjoint,
face-conforming tetrahedra $\D_k$ with boundary conditions
\[ 
  u|_{\Gamma_i} (x, t) = g_i (u (x, t), x, t), \hspace{2em} i = 1, \ldots, b,
\]
at inflow boundaries $\biguplus \Gamma_i \subseteq \partial \Omega$. 
As stated, we will assume the flux function $F$ to be linear.  
We find a weak form of (\ref{eq:claw}) on each element $\D_k$:
\begin{align*} 
  0 & = \int_{\D_k} u_t \varphi + [\nabla \cdot F (u)] \varphi \mathd x\\ 
  & = \int_{\D_k} u_t \varphi - F (u) \cdot \nabla \varphi \mathd x
  + \int_{\partial \D_k} ( \hat{n} \cdot F)^{\ast} \varphi \mathd S_x, 
\end{align*}
where $\varphi$ is a test function, and $( \hat{n} \cdot F)^{\ast}$ is a
suitably chosen numerical flux in the unit normal direction $\hat{n}$.
Following \cite{hesthaven_nodal_2007}, we find a `strong'-DG form of
this system as
\begin{equation}
  0 = \int_{\D_k} u_t \varphi + [\nabla \cdot F (u)] \varphi \mathd x-
  \int_{\partial \D_k} [ \hat{n} \cdot F - ( \hat{n} \cdot F)^{\ast}] 
  \varphi \mathd S_x.
  \label{eq:strong-dg}
\end{equation}
We seek to find a numerical vector solution $u^k \assign u_N|_{\D_k}$ from the
space $P_N^n (\D_k)$ of local polynomials of maximum total degree $N$ on each
element.  We choose the scalar test function $\varphi \in P_N (\D_k)$ from the
same space and represent both by expansion in a basis of $N_p\assign
\operatorname{dim} P_N(\D_k)$ Lagrange polynomials $l_i$ with respect to a set
of interpolation nodes \cite{warburton_explicit_2006}. We define the mass,
stiffness, differentiation, and face mass matrices
\begin{subequations} 
  \label{eq:dg-global-matrices} 
  \begin{align} 
    M_{i j}^k & \assign \int_{\D_k} l_i l_j \mathd x, \\ 
    S_{i j}^{k, \partial\nu} & \assign \int_{\D_k} l_i \partial_{x_{\nu}} l_j \mathd x,\\ 
    D^{k, \partial\nu} & \assign (M^k)^{- 1} S^{k, \partial\nu},\\ 
    M_{i j}^{k, A} & \assign \int_{A \subset \partial \D_k} l_i l_j \mathd S_x.
  \end{align} 
\end{subequations}
Using these matrices, we rewrite \eqref{eq:strong-dg} as
\begin{align} 
  0 & =  M^k \partial_t u^k 
  + \sum_{\nu} S^{k, \partial_{\nu}} [F(u^k)]
  - \sum_{F \subset \partial \D_k} M^{k, A} [
  \hat{n} \cdot F - ( \hat{n} \cdot F)^{\ast}], \notag \\
  \label{eq:semidiscrete-dg} 
  \partial_t u^k & = 
  - \sum_{\nu} D^{k, \partial_{\nu}} [F(u^k)]
  + L^k [ \hat{n} \cdot F 
  - ( \hat{n} \cdot F)^{\ast}] |_{A \subset \partial \D_k}.
\end{align}
The matrix $L^k$ used in \eqref{eq:semidiscrete-dg} deserves a little
more explanation. It acts on vectors of the shape
$[u^k|_{A_1},\dots,u^k|_{A_4}]^T$, where $u^k|_{A_i}$ is the vector of
facial degrees of freedom on face $i$. For these vectors, $L^k$
combines the effect of applying each face's mass matrix, embedding the
resulting facial values back into a volume vector, and applying the
inverse volume mass matrix.  Since it ``lifts'' facial contributions
to volume contributions, it is called the \emph{lifting matrix}. Its
construction is shown in Figure \ref{fig:lifting-matrix}. 

It deserves explicit mention at this point that the left
multiplication by the inverse of the mass matrix that yields the
explicit semidiscrete scheme \eqref{eq:semidiscrete-dg} is an
element-wise operation and therefore feasible without global
communication. This strongly distinguishes DG from other finite
element methods. It enables the use of explicit (e.g., Runge-Kutta)
time stepping and greatly simplifies parallel implementation efforts
such as this one.

\subsection{Implementing DG}

\label{ssec:implement-dg}

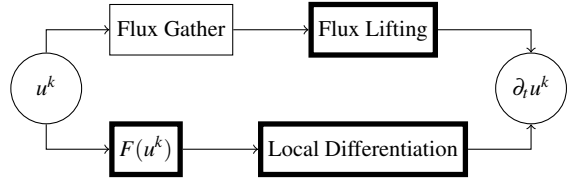
\begin{figure}
  \centering
  \begin{tikzpicture}[
    txt/.style={text height=1.5ex, text depth=0.25ex},
    operation/.style={rectangle,draw,minimum height=5ex,txt},
    localop/.style={operation,line width=2pt,txt},
    data/.style={circle,draw,minimum size=7ex,txt},
    ]

    \node [data] (state) { $u^k$ } ;
    \node [operation] (gather) [above right=0.1cm and 0.5cm of state] { Flux Gather } ;
    \node [localop] (lift) [right=1cm of gather] { Flux Lifting } ;
    \node [localop] (fu) [below right=0.1cm and 0.5cm of state] { $F(u^k)$ } ;
    \node [localop] (diff) [right=1cm of fu] { Local Differentiation} ;
    \node [data] (rhs) [above right=0.1cm and 0.5cm of diff] { $\partial_t u^k$ } ;
    \draw [->] (state) |- (gather) ;
    \draw [->] (gather) |- (lift) ;
    \draw [->] (lift) -| (rhs) ;

    \draw [->] (state) |- (fu) ;
    \draw [->] (fu) |- (diff) ;
    \draw [->] (diff) -| (rhs) ;
  \end{tikzpicture}
  \caption[Decomposition of a DG operator into subtasks.]
  {Decomposition of a DG operator into subtasks. Element-local
  operations are highlighted with a bold outline.}
  \label{fig:dg-subtasks}
\end{figure}

DG decomposes very naturally into four stages, as visualized in Figure
\ref{fig:dg-subtasks}.  This clean decomposition of tasks stems from
the fact that the discrete DG operator \eqref{eq:semidiscrete-dg} has
two additive terms, one involving an element volume integral, the
other an element surface integral.  The surface integral term then
decomposes further into a `gather' stage that computes the term
\begin{equation} 
  [ \hat{n} \cdot F(u_N^-) - ( \hat{n} \cdot F)^{\ast}(u_N^-, u_N^+)]|_{A \subset
  \partial \D_k},
  \label{eq:num-flux} 
\end{equation}
and a subsequent lifting stage. The notation $u_N^-$ indicates the
value of $u_N$ on the face $A$ of element $\D_k$, $u_N^+$ the value of
$u_N$ on the face opposite to $A$.

As is apparent from the use of a Lagrange basis, we employ a
\emph{nodal} version of DG, in which the stored degrees of freedom
(``\emph{DOF}s'') represent the values of $u_N$ at a set of
interpolation nodes. This representation allows us to find the facial
values used in \eqref{eq:num-flux} by picking the facial nodes from
the volume field. (This contrasts with a \emph{modal} implementation
in which DOFs represent expansion coefficients in a non-Lagrange
basis.  Finding the facial information to compute \eqref{eq:num-flux}
requires a different approach in these schemes.)

Observe that most of DG's stages are \emph{element-local} in the sense
that they do not use information from neighboring elements. Moreover,
these local operations are often efficiently represented by a dense
matrix-vector multiplication on each element.

It is worth noting that since simplicial elements only require affine
transformations $\Psi_k$ from reference to global element, the global
matrices can easily be expressed in terms of reference matrices that
are the same for each element, combined with scaling or linear
combination, for example
\begin{subequations}
\label{eq:dg-reference-matrices}
\begin{align}
  M_{i j}^k & = 
  \underbrace{\left|\det \frac{\mathd \Psi_k}{\mathd r}\right|}_{J_k\assign}
  \underbrace{\int_{\mathsf{I}} l_i l_j \mathd x}_{M_{ij}\assign},\\
  S_{i j}^{k, \partial \nu} & = J_k \sum_{\mu} \frac{\partial \Psi_{\nu}}{\partial r_{\mu}}
  \underbrace{\int_{\mathsf{I}} l_i \partial_{r_{\mu}} l_j \mathd x}_{S_{ij}^{\partial \mu}\assign},
\end{align}
\end{subequations}
where $\mathsf{I}=\Psi_k^{-1}(\D_k)$ is a reference element. We define
the remaining reference matrices $D$, $M^A$, and $L$ in an analogous
fashion.

\section{GPU-DG: Motivation and Challenges}
\label{sec:gpu-dg}

As we begin our study of bringing DG methods onto GPU-like
architectures, we should first establish what we intend to achieve in
doing so. Our main motivation is a gain in performance available from
a desk-side workstations. We believe that the amount of computing
power easily available to an engineer often determines the amount of
computing power used in a given engineering challenge. Remote
resources such as big clusters provide large amounts of power quite
readily, but their use also implies a complexity burden in management,
cost, and access. Nonetheless, good performance on clusters and large
machines is clearly a secondary goal. Next, we would like to be able
to apply the technology under discussion to a wide range of partial
differential equations. While DG methods are designed for and best
suited to hyperbolic PDEs, there is no conceptual restriction to this
type of PDE--and our GPU-DG technology is not restricted in this way,
either. A tertiary goal of ours is to make the technology not just
worthwhile on a desk-side workstation, but also simple enough to apply
that an engineer can easily manage his or her own computations. While
this is partially a software design issue beyond the scope of this
article, some prerequisites at the GPU computation level must be met
to accommodate the desired ease of use. In particular, no knowledge of
GPU computing is necessary to manage a computation.

The discontinuous Galerkin method further allows considerable user
choice at the level of the reference discretization. It is not
practical to support \emph{all} such choices, and thus we introduce
the following (fairly non-restrictive) stipulations:
\begin{itemize}
  \item We will specialize to straight-sided simplices, as we perceive
  the required volume mesh generation machinery to be the most mature
  for this type of element. The restriction to straight-sidedness is
  comparatively easy to lift \citep{warburton_low-storage_2010}.
  \item We will optimize for (but not specialize to) the three-dimensional 
  case, i.e. tetrahedral elements, as it bears both the most relevance
  to application problems and the greatest computational complexity.
  \item We will further optimize for ``medium'' order ($N=3\dots5$)
  polynomial spaces, as those maintain the DG time step restriction
  ($\Delta t\sim \Delta x/N^2$, see \citep{hesthaven_nodal_2007})
  at a reasonable level.
\end{itemize}

Next, we consider what possible obstacles our effort to bring DG to
the GPU may face. Perhaps the first challenge that comes to mind is
that of data movement. As in any matrix-product-type workload, there
is much data reuse in DG, but matrices grow rapidly as $N$ increases.
In terms of data reuse, that is good--however it does compete with the
limited size of on-chip memories that are needed to realize the
possible reuse. Further, while on-chip memories are growing at a
moderate pace and management of these memories becomes more automatic,
it should be noted that even CPU-based matrix-matrix multiplies
benefit from explicit management of their L1 caches
\citep{bilmes_optimizing_1997,whaley_automated_2001} in matrix-based
workloads. Strongly interwoven with the challenge of having to manage
on-chip memories is that of accommodating hardware granularities, such
as memory sizes, memory bus widths, SIMD widths, workgroup sizes, and
so on. In addition, discontinuous Galerkin methods have their own
preferred granularities, such as the number of degrees of freedom in
each element, the number of dimensions, or the number of degrees of
freedom on an element's face. Unfortunately, while machine-related
granularities have a tendency to be powers of two, the exact opposite
is true of those related to the numerical method. Another challenge
posed by GPU computing is the necessity to map the computation onto a
two-level, grid-based parallel execution structure, with the first
level corresponding to parallelization across cores, and the second to
parallelization across SIMD lanes within a core. While a coarse-grain
structure may often be immediate, various finer details of this choice
require careful tuning.

\begin{figure}
  \begin{tikzpicture}
    [font=\sffamily\scriptsize,scale=0.7,
    ]
    \foreach\i in {2,1,0}
      \node [anchor=south west,draw,fill=yellow,
        text width=2cm,minimum height=2cm,
        text centered,anchor=east]
        at (-1,0,-\i*0.5) (locmat)
        { Local Templated Derivative Matrices };

    \draw [|<->|]
      ($ (locmat.south east) + (0,-0.15) $)
      -- ($ (locmat.south west) + (0,-0.15) $) 
      node [pos=0.5,below=0.1cm] {$N_p$}
      ;
    \draw [|<->|]
      ($ (locmat.south west) + (-0.15,0) $)
      -- ($ (locmat.north west) + (-0.15,0) $) 
      node [pos=0.5,left=0.1cm] {$N_p$}
      ;

    \foreach \i in {0} {
      \draw [fill=green!50,opacity=1-\i*0.16] (0,-1.5,-\i*0.5) rectangle ++(3,3) ;
    }
    \foreach \x in {0,0.2,...,3}
      \draw [opacity=0.3] (\x,-1.5) -- ++(0,3) ;

    \draw [|<->|]
      ($ (0,1.5) + (0,0.15) $)
      -- ($ (3,1.5) + (0,0.15) $) 
      node [pos=0.5,above=0.1cm] {$K$}
      ;

    \draw [|<->|]
      ($ (3,-1.5) + (0.15,0) $)
      -- ($ (3,1.5) + (0.15,0) $) 
      node [pos=0.5,right=0.1cm] {$N_p$}
      ;

    \node [anchor=center,fill=green!50, text centered]
      at (1.5,0)
      { Field Data } ;

    \draw [fill=blue!50] (0,-2) rectangle ++(3,-0.75) ;
    \foreach \x in {0,0.2,...,3}
      \draw [opacity=0.3] (\x,-1.5) -- ++(0,3) ;

    \node [anchor=center,fill=blue!50]
      at (1.5,-2-0.75*0.5)
      { Geom. Factors } ;
  \end{tikzpicture}
  \begin{tikzpicture}
    [font=\sffamily\scriptsize,scale=0.7
    ]
      \node [draw,fill=yellow,
        text width=3cm,minimum height=2cm,
        text centered,anchor=east]
        at (-0.5,0) (locmat)
        { Local (Templated) Lifting Matrix };

      \draw [|<->|]
        ($ (locmat.south east) + (0,-0.15) $)
        -- ($ (locmat.south west) + (0,-0.15) $) 
        node [pos=0.5,below=0.1cm] {$N_fN_{fp}$}
        ;
      \draw [|<->|]
        ($ (locmat.south west) + (-0.15,0) $)
        -- ($ (locmat.north west) + (-0.15,0) $) 
        node [pos=0.5,left=0.1cm] {$N_p$}
        ;

    \draw [fill=green!50,opacity=1] (0,-1.5) rectangle ++(3,3) ;
    \foreach \x in {0,0.2,...,3}
      \draw [opacity=0.3] (\x,-1.5) -- +(0,3) ;

    \draw [|<->|]
      ($ (0,1.5) + (0,0.15) $)
      -- ($ (3,1.5) + (0,0.15) $) 
      node [pos=0.5,above=0.1cm] {$K$}
      ;

    \draw [|<->|]
      ($ (3,-1.5) + (0.15,0) $)
      -- ($ (3,1.5) + (0.15,0) $) 
      node [pos=0.5,right=0.1cm] {$N_fN_{fp}$}
      ;

    \node [anchor=center,fill=green!50, text centered]
      at (1.5,0)
      (field-data)
      { Facial Field Data } ;

    \begin{scope}[yshift=-1.75cm]
      \draw [fill=blue!50] (0,0) rectangle ++(3,-0.5) ;
      \draw [xstep=0.2cm,ystep=5cm,opacity=0.3] (0,0) grid ++(3,-0.5) ;

      \node [anchor=center,fill=white, fill=blue!50,inner ysep=0]
        at (1.5,-0.5*0.5)
        { (Inv.) Jacobians} ;
    \end{scope}
  \end{tikzpicture}
  \caption{Workload size characterization for element-local linear operators.
  Left: Element-local differentiation. Right: Lifting from flux values
  along element faces into volume data. In each case, matrix sizes are
  given in terms of the quantities of Section \ref{sec:dg-overview}.}
  \label{fig:el-local-ops}
\end{figure}
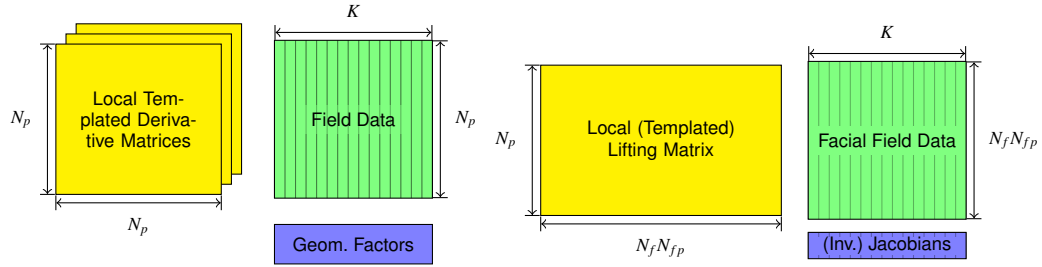

We will now discuss how these challenges are met by our approach,
through a few representative examples, beginning with the question of
data movement for element-local linear operators such as lifting and
elementwise differentiation. Figure \ref{fig:el-local-ops} illustrates
the type and size of data that these procedures operate on. The figure
also makes it obvious that while the method primarily relies on
matrix-vector products, it is profitable to view the field vectors in
aggregate as a matrix, thereby giving rise to a matrix-matrix
computation, albeit with very off-balance matrix dimensions.  An
obvious first approach would be to use vendor-supplied BLAS matrix
libraries for such a task, however it turns out that these are often
tuned for large, square matrices and rarely deal well with the matrix
sizes occurring in DG. One is therefore left to build a home-grown
algorithm. Given that, depending on the local polynomial order $N$,
only a limited amount of this data can fit onto the chip, the
implementer is faced with a decision of which data to store locally
and which data to stream onto the chip. In particular, one might
consider the following alternatives:
\begin{itemize}
  \item Store the matrices, stream the vector data. This seems like an
  obvious choice--however the matrices are often too large, and vector
  data is much more easily partitioned.
  \item Store part of a matrix. This complicates the access logic, but
  can often profitably be done--especially by using row-wise
  partitions.
  \item Store only field data. If streaming of the matrix is achieved
  through a cached data path, this can also be an attractive option.
  \item Store parts of the matrix and the field vector. While this
  could, in theory, provide the best balance of data reuse, we were
  unable to turns this approach into competitive code.
\end{itemize}
This is a choice that an implementer needs to make, however we have
found no universally valid heuristic that might provide guidance on
which alternative to choose, especially given that the optimality of
each option strongly depends on the hardware being used.

\begin{figure}
  \begin{center}
    \sffamily\scriptsize
    \def\workunit#1{
      \draw [draw,fill=green] #1 ++(0,0) rectangle +(0.15,-0.15) ;
      \draw [draw,fill=red] #1 ++(0,-0.15) rectangle +(0.15,-0.15) ;
      \draw [draw,fill=yellow] #1 ++(0,-0.3) rectangle +(0.15,-0.15) ;
      \draw [draw,fill=blue] #1 ++(0,-0.45) rectangle +(0.15,-0.15) ;
    }
    \newcommand\drawaxes{
      \draw [->] (-0.2,0.5) -- +(1,0) node [pos=1,right=0.1cm] {Work Item};
      \draw [->] (-0.2,0.5) -- +(0,-1) node [pos=1,below=0.1cm] {$t$};
    }
    \tikzstyle{wuprep}=[draw,fill=cyan]
    \textbf{Per Block:} $K_L$ element-local mat.mult. + matrix load

    \smallskip
    \begin{tikzpicture}
      \node at (0,0.15) [anchor=south west,
        wuprep,minimum width=6.15cm,inner sep=0]
        {Preparation};
      \foreach \x in {0,0.3,...,6.1} { \workunit{(\x,0)}  }
    \end{tikzpicture}
    \smallskip


    \medskip
    \begin{tabular}{p{0.25\textwidth}p{0.25\textwidth}p{0.25\textwidth}}
    $w_s$: in sequence

    \begin{tikzpicture}
      \drawaxes
      \workunit{(0,0)}
      \workunit{(0,-0.7)}
      \workunit{(0,-1.4)}
      \draw [wuprep] (0,0.15) rectangle +(0.15,0.15) ;
    \end{tikzpicture}
    &
    $w_i$: ``inline-parallel''

    \begin{tikzpicture}
      \drawaxes
      \draw [fill=green] (0,0) rectangle +(0.15,-0.45) ;
      \draw [fill=red] (0,-0.45) rectangle +(0.15,-0.45) ;
      \draw [fill=yellow] (0,-0.9) rectangle +(0.15,-0.45) ;
      \draw [fill=blue] (0,-1.35) rectangle +(0.15,-0.45) ;

      \draw [draw,ystep=0.14999cm] (0,-1.8) grid +(0.15,1.8) ;

      \draw [wuprep] (0,0.15) rectangle +(0.15,0.15) ;
    \end{tikzpicture}
    &
    $w_p$: in parallel

    \begin{tikzpicture}
      \drawaxes
      \workunit{(0,0)}
      \workunit{(0.3,0)}
      \workunit{(0.6,0)}
      \draw [wuprep] (0,0.15) rectangle +(0.15,0.15) ;
      \draw [wuprep] (0.3,0.15) rectangle +(0.15,0.15) ;
      \draw [wuprep] (0.6,0.15) rectangle +(0.15,0.15) ;
    \end{tikzpicture}
    \\
    (amortize preparation)
    &
    (exploit register space)
    \end{tabular}
  \end{center}
  \caption{Choices for the amount of work done by a workgroup in an
  element-local (differentiation, lift) operation.}
  \label{fig:el-local-workgroup}
\end{figure}
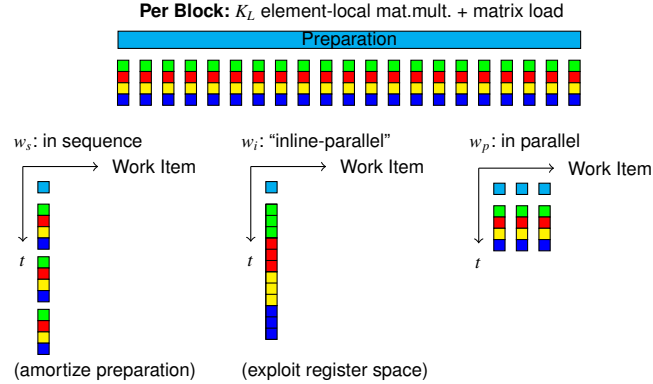

For the same workload of elementwise local differentiation and
lifting, there is also the question of which work decomposition to
use, where the work decomposition is given (in vendor-neutral OpenCL
terminology) by the number of workgroups and their sizes. Each
of these quantities can further be decomposed into a three-component
vector. Order in this three-component vector matters, as it determines
which work items execute memory accesses at the same time, and which
branches may require serialization. 

Abstractly, the workload under consideration consists of an (optional)
preparatory step that preloads matrix data into on-chip memory,
followed by dot products for each matrix row and all the columns
(field vectors). The most immediate choice would be to have each
workgroup deal with one such matrix-vector product, leading to a
one-to-one mapping between workgroups and DG elements. While this is
certainly straightforward, it has a number of drawbacks. For the
polynomial orders $N$ targeted in this work, these workgroup sizes are
unable to fill the (32- or 64-)wide SIMD architectures exhibited by
today's GPUs--at least not efficiently, and not without leaving unused
'gaps' in the SIMD vector. Further,  one also typically adds padding
to conform to a device's memory alignment, and this choice leads to a
maximum number of gaps in the data, thereby wasting a considerable
amount of (typically precious) GPU memory. In addition to that, if one
workgroup only performs one matrix-vector product, any preparation
steps would be poorly amortized.

Irrespective of more advanced blocking options (as described in
\citep{kloeckner_nodal_2009}), there are three basic, orthogonal (i.e.
arbitrarily combinable) possibilities of remedying this situation,
outlined in Figure \ref{fig:el-local-workgroup}. Two of these choices
are entirely obvious, the third slightly less so. The obvious choices
include letting a workgroup do more things \emph{in sequence} and
\emph{in parallel}. The former of these leads to better amortization
of preparation steps, while the latter does that, and in addition
increases utilization of parallel processing resources. The third
choice, called \emph{in-line parallel} in Figure
\ref{fig:el-local-workgroup}, occupies a middle-ground between the
two by accomplishing a number of dot products along with each other
within a single work item. This exploits the fact that, in order for
the matrix to be operated on, its components must be resident within
the GPU's register file--but once they are there, it is economical to
use them not just once, but multiple times. All of these strategies
are specific forms of \emph{work item coarsening}. How many
elements are worked on in each of these fashions is captured by the
numbers $w_s$, $w_i$ and $w_p$.

Obviously, regardless of the choices for these numbers, the same
amount of work is begin done--it is just the partitioning that
differs. Nonetheless, in Section \ref{sec:results}, we will observe
fairly significant performance differences between such partitionings.

\begin{figure}
  \centering
  \def\myxscale{0.14cm}
  \def\myyscale{0.2cm}
  \def\drawmeshclump#1{{
    \def\size{#1}
    \pgfmathsetmacro{\smone}{\size-1}
    \foreach \i in {1,...,\smone}
    {
      \pgfmathsetmacro{\imone}{\i-1}
      \foreach \j in {0,...,\imone}
      {
        \filldraw (-\i+2*\j,\i) -- ++(2,0) -- ++(-1,-1) --cycle;
        \filldraw (-\i+2*\j,\i) ++(1,-1) -- ++(2,0) -- ++(-1,1) --cycle;
      }
      \filldraw (-\i+2*\i,\i) -- ++(2,0) -- ++(-1,-1) --cycle;
    }
    \filldraw (0,0) -- ++(2,0) -- ++(-1,-1) --cycle;
  }}

  \def\drawmesh#1#2{{
    \def\size{#1}
    \def\clumpsize{#2}
    \pgfmathsetmacro{\offset}{\clumpsize+1}
    \pgfmathsetmacro{\smone}{\size-1}
    \foreach \i in {1,...,\smone}
    {
      \pgfmathsetmacro{\imone}{\i-1}
      \foreach \j in {0,...,\i}
      {
        \begin{scope}
        [xshift=(-\i+2*\j)*\offset*\myxscale,
        yshift=\offset*\i*\myyscale]
          \drawmeshclump{\clumpsize}
        \end{scope}

        \begin{scope}
        [
        xshift=(-\i+2*\j+1)*\offset*\myxscale,
        yshift=(-2+2*\clumpsize+0.7+\offset*(\i-1))*\myyscale,
        ]
          \begin{scope}[yscale=-1]
            \drawmeshclump{\clumpsize}
          \end{scope}
        \end{scope}
      }
      \drawmeshclump{\clumpsize}
    }
  }}
  \begin{tikzpicture}
  [x=\myxscale,y=\myyscale,thick,fill=yellow]
    \drawmesh 4 2
    \begin{scope}[xshift=4cm]
     \drawmesh 3 4
    \end{scope}
  \end{tikzpicture}
  \caption{Multiple granularities for inter-element flux computation.
  Obviously, larger blocks lead to more data reuse as fewer face pairs
  are split.}
  \label{fig:flux-granularity}
\end{figure}
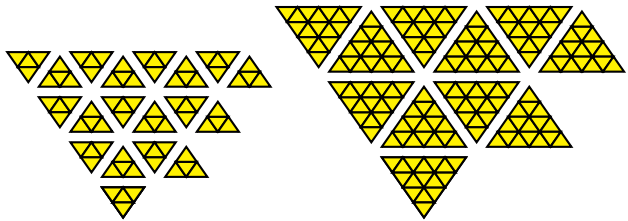


We have just seen that a question of granularities arises even in a
simple situation like that of the element-local operations. There is an even
more important concern of this nature in the only inter-element
communication operation within DG, the computation of surface fluxes.
Since the computation of each surface flux refers to data from two
opposite element faces, there is definite savings potential if data
for a number of such faces is brought onto the chip at the same time
and reused. Obviously, this leads to a decrease in the amount of
parallelism available, but for large enough problems (which are the
main driver for the application of GPU technology), this becomes a
non-issue. The amount of parallelism is however limited by two sets of
data that need to be fit onto the chip, namely the metadata indicating
which faces with what geometry data need to be processed, and the
output buffer used to write vectors of face data that can then be
processed in the lifting stage of the computation. Both of these could
theoretically be accomplished in streaming mode without on-chip
storage, however we have found that buffering them improves
performance measurably. Once a granularity has been found that
suitably balances these factors with data reuse, the computational
mesh needs to be partitioned in a way that maximizes the number of
interior faces in each partition. Fortunately, we have found that
performance is somewhat insensitive to the absolute quality of this
partition, and a simple greedy algorithm, as outlined in
\citep{kloeckner_nodal_2009}, suffices.

\begin{figure}
  \sidecaption
  \begin{tikzpicture}[scale=0.55,font=\scriptsize]
    \foreach \bottom in {0, -1, -2}
    {
      \draw (0,\bottom) rectangle +(6.4,1) ;
      \draw [fill=green!40] (0,\bottom) rectangle +(2,1) ;
      \draw [fill=green!40] (2,\bottom) rectangle +(2,1) ;
      \draw [fill=green!40] (4,\bottom) rectangle +(2,1) ;
    }
    \foreach \i in {0,0.1,...,6.4}
    { \draw (\i,1) -- (\i,0.9); }
    \foreach \x in {1,3,5}
    {
      \node at (\x,-1.5) {Element} ;
      \node at (\x,-0.5) {Element} ;
      \node at (\x,0.4) {\dots} ;
    }
    \node (padding) at (6,2.3) {Padding} ;
    \draw [->] (padding) -- (6.2,0.5) ;
    \draw [|<->|] (0,-2.5) -- +(2,0)
      node [pos=0.5,anchor=north] {$N_p$} ;
    \draw [|<->|] (0,-2.25) -- +(6,0) 
      node [pos=0.5,anchor=north] {$K_M N_p$} ;
    \node [left] at (-0.2,0.5) { 128 };
    \node [left] at (-0.2,-0.5) { 64 };
    \node [left] at (-0.2,-1.5) { 0 };
  \end{tikzpicture}
  \caption{Element storage in ``microblock'' format as described in
  the text. An small, integer number of elements is followed by enough
  padding to satisfy device alignment requirements. Other computation
  granularities are specified as integer numbers of microblocks.}
  \label{fig:microblock}
\end{figure}
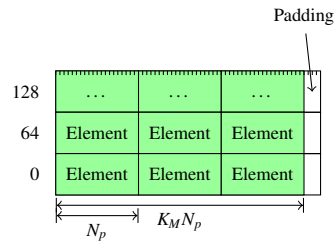

Overall, we have seen a few examples of computations requiring that
the implementer select a granularity entirely unrelated to the
computation itself. Each of these granularities is bound to want to
manifest itself somehow in the in-memory data storage format, likely
through coalescing/alignment concerns. On the other hand, it is
\emph{not} likely that a single data storage format can satisfy
\emph{all} restrictions of \emph{all} parts of the computation. A
compromise therefore needs to be made. In calling the granularities of
each of the computations ``\emph{blocks}'' (related to the Nvidia term
for workgroups), we arrived at the idea of an intermediate granularity
consisting of an integer number of elements and just big enough to
satisfy the device's basic alignment preference, but not necessarily
conforming to any particular computation. This would then be called a
``\emph{microblock}'' (illustrated in Figure \ref{fig:microblock}),
and we would demand that all the actual computation granularities be
integer multiples of a microblock. A similar technique was
independently discovered in \citep{filipovic_medium-grained_2010}.
Seemingly, this just introduces yet another semi-arbitrary number to
be chosen before the computation can begin, but nonetheless its
introduction does some good by relieving the tension over the data
storage format between different parts of the computation.

As we conclude our overview of a few of the challenges of bringing
discontinuous Galerkin methods onto the GPU, we observe that there is
a common theme uniting many of them--the answers are strongly
hardware-dependent. This has a number of important consequences:
\begin{itemize}
  \item The questions themselves are difficult to answer. Modern
  processor hardware tends to be very complicated, with many clock
  domains, bandwidth figures, possibilities for resource contention,
  and so on.
  \item Published information on hardware provides insufficient
  heuristics to make well-founded decisions on any of these.
  \item Even if a good answer to these questions existed, then it
  would not necessarily have any lasting value. Software tends to have
  a much longer shelf life than hardware, as new hardware revisions
  with programmer-visible changes to microarchitecture (in both the
  GPU and CPU markets) appear at a rate of about one every two years.
  Some things (such as the OpenCL programming model) are expected to
  be durable for at least some time, but the fine features determining
  tuning decisions such as those outlined above are subject to
  frequent change.
\end{itemize}
One obvious solution to this tuning dilemma stems from the realization
that computer cycles are cheap--and that it is thus reasonable to let
a computer help as much as possible in solving these challenges. If
that means letting the machine try out a large number of possible
combinations of parameter settings, that is fine--computer time is
less expensive than human time, and this trend will almost certainly
continue.  Furthermore, this shifts two aspects of GPU programming in
the right direction. First, it shifts the programmer's role from
caring about tuning results to coming up with tuning ideas--and letting
the computer determine to what extent those are effective. Second, it
decreases the amount of detailed hardware knowledge necessary to come
up with a high-performance program. Arguably, both of these represent
steps in the right direction. In the next section, we will present
ideas on the concrete implementation of \emph{automated tuning}.

\section{Run-Time Code Generation}
\label{sec:rtcg}
The capability to do automated tuning, i.e. to do an automated
benchmark of a large number of variants of a program turns out to be a
special case of a much more general facility--that of \emph{Run-Time
Code Generation} (``RTCG'').

This phrase has two parts, `code generation' and `run-time'. In
itself, the \emph{generation} of source code is merely a text
processing task, and most modern languages, which most modern
languages are more than capable of.  What is being discussed here is
thus not the actual generation of the code (which is just a piece of
ASCII text), but rather the ability to compile and run this code
in-process, at \emph{run-time}.

GPU programming environments such Nvidia's CUDA ``runtime'' interface
make this difficult because they insist that all code be compiled
ahead of time and into one final binary . In such a setting, all
possible tuning variants must already be precompiled into the
application binary. This restriction can of course be worked around
using dynamic linking and/or shell scripting, but none of these lead
to particularly elegant or robust solutions.

We aim to demonstrate in this article that \emph{scripting languages}
make a very hospitable environment for run-time code generation,
especially when using interfaces such as OpenCL or the Nvidia CUDA
``driver'' interface which facilitate RTCG more easily.  Scripting
languages usually have no need for a user- or developer-visible
compilation step, and thus everything they do is, by definition, done
at run time.

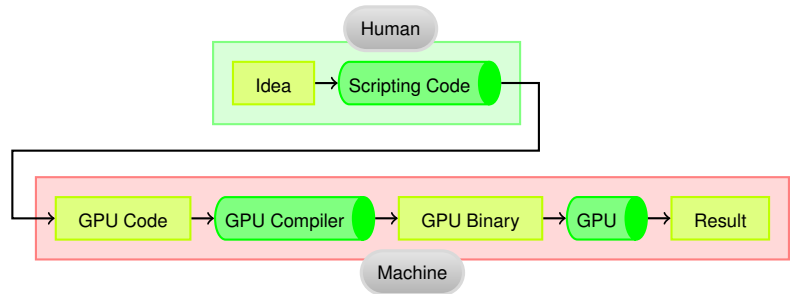
\begin{figure}
  \centering
  \beginpgfgraphicnamed{gpu-code-generation}
  \begin{tikzpicture}[
    font=\sffamily\scriptsize,
    actor/.style={cylinder,
      fill=none,draw=green,thick,
      cylinder uses custom fill=true,
      cylinder end fill=green,
      cylinder body fill=green!50,
      shape aspect=.5,
      text height=1em,
      text depth=0.5ex,
      text centered,
      },
    object/.style={
      rectangle,
      minimum height=3.5ex,
      inner xsep=3mm,
      fill=lime!50,
      draw=lime,
      text height=1em,
      text depth=0.5ex,
      thick,
      },
    hlbox/.style={
        rectangle,fill=#1!30,draw=#1,opacity=0.5,thick
    },
    button/.style={
      rounded rectangle,
      bottom color=gray!60,
      top color=gray!20,
      inner sep=2mm,
      draw=gray!30,very thick,
      },
    ]
    \node [object] (idea) at (0,0) {Idea} ;
    \node [actor,right=0.3 of idea] (py) {Scripting Code} ;
    \draw [->,thick] (idea) -- (py) ;

    \node [object] at (-2,-1.8) (cu) {GPU Code} ;
    \draw [->,thick] (py.east) -| ++(0.5,-0.9)  -| ++(-7,0) |- (cu.west) ;
    \node [actor,right=0.3 of cu] (nvcc) {GPU Compiler} ;
    \draw [->,thick] (cu) -- (nvcc) ;
    \node [object,right=0.3 of nvcc] (cubin) {GPU Binary} ;
    \draw [->,thick] (nvcc) -- (cubin) ;
    \node [actor,right=0.3 of cubin] (gpu) {GPU} ;
    \draw [->,thick] (cubin) -- (gpu) ;
    \node [object,right=0.3 of gpu] (result) {Result} ;
    \draw [->,thick] (gpu) -- (result) ;

    \begin{pgfonlayer}{background}
      \node [hlbox=green,fit=(idea) (py),inner sep=2.5mm] (human) {};
      \node at (human.north) [yshift=-0.15cm,anchor=south west,button] {Human} ;
    \end{pgfonlayer}

    \begin{pgfonlayer}{background}
      \node [hlbox=red,fit=(cu) (result),inner sep=2.5mm] (machine) {};
      \node [below=-0.15cm of machine,button] {Machine} ;
    \end{pgfonlayer}

  \end{tikzpicture}
  \endpgfgraphicnamed
  \caption{Operating principle of GPU code generation.}
  \label{fig:gpu-code-generation}
\end{figure}

Even beyond what was discussed so far, there are many good reasons
to ask for the ability to do run-time code generation:
\begin{itemize}
  \item \emph{Automated Tuning}, as discussed. (This is also done,
  although in a variety of ways accommodating ahead-of-time
  compilation, by packages such as ATLAS
  \citep{whaley_automated_2001}, FFTW \citep{frigo_2005}, or PHiPAC
  \citep{bilmes_optimizing_1997}.)
  \item The ability to vary \emph{data types} at run time. This might
  include the ability to run in double or single precision, with
  complex- or real-valued data, or even more complicated variants,
  such as interval arithmetic. Templates (in C++) partially address
  this need in the ahead-of-time-compiled world, however, they also
  incur the overhead of having to compile each possible variant into
  the executed binary, as there is no possibility for compilation at
  run time.
  \item From the perspective of a library writer, another attractive
  possibility opened up by RTCG is the possibility to \emph{specialize
  code for a user-given problem}. While many rather complicated
  systems involving C++ metaprogramming strive to achieve this goal,
  they cannot match the simplicity (and performance) of textually
  pasting a chunk of purpose-specific C code into an overall code
  framework. Also note that this benefit is really not specific to
  library writers at all.  At some level, every programmer strives to
  write code that is general and covers a wide variety of use cases.
  RTCG opens up a very simple and high-performance avenue towards this
  goal.
  \item Lastly, it should be observed that \emph{constants faster 
  than variables}. This can be easily understood from the standpoint
  of \emph{register pressure}--where space in the register file is just one
  of many resources that are scarce in a GPU, and less contention
  means that some trade-off does not need to be made, which usually
  results in higher performance. Another specific example of this 
  is \emph{loop unrolling}. Loops with unknown trip counts necessarily
  come with fixed overhead in the form of end-of-loop tests and
  branching instructions, in addition to loop-related state being kept
  in the register file. If the loop trip count is known at
  run-time, then this overhead is easily done away with.
\end{itemize}
All of these arguments in favor of RTCG rest on a simple fact:
\emph{More information is available to a code generator and compiler
at run time than at any time before that.} And unsurprisingly, the
more information is available to the code generator and the compiler,
the better the code it is able to generate. Also observe that in this
picture, the code generator and the compiler start to merge together
conceptually, and the representation in which they exchange data
(often a variant of C, for now) moves towards being an implementation
detail. This is a good thing, as it makes it expedient for programmers
to develop representations that best serve their application.
Interfaces like CorePy \citep{mueller_corepy_2007} and LLVM
\citep{lattner_llvm_2004} demonstrate that C is not the only possible
intermediate representation.

As we discuss the advantages of RTCG, we should likely also mention
the (in our opinion few and minor) disadvantages. First, RTCG
obviously adds more moving parts (such as a compiler and a
just-in-time execution environment) to a program, which introduces
more possible sources of issues. Second, as generated code must be
compiled, there is often a noticeable delay before a piece of code is
first executed. However, caching and parallel compilation are
effective remedies for this.

Despite these perceived drawbacks, the creators of the OpenCL
specification seem to agree with our point of view and have made RTCG
a standard part of the OpenCL interface--which, in our opinion, is one
of the most interesting contributions OpenCL makes to the high-performance
computing arena.  When OpenCL is compared to CUDA, one drawback that
is often cited is OpenCL's lack of support for C++ templates. This is
a moot point, in our opinion, as RTCG is strictly more powerful than
C++ templates.

Next, we would like to continue to argue that RTCG is most effective
when practiced from a scripting language. Scripting languages are in
many ways polar opposites to GPUs. GPUs are highly parallel, subject
to hardware subtleties, and designed for maximum throughput. On the
other hand, scripting languages (such as Python
\citep{vanrossum_python_1994}) favor ease of use over computational
speed, are largely hardware-agnostic, and do not generally emphasize
parallelism. We have created two packages, PyOpenCL and PyCUDA
\citep{kloeckner_pycuda_2009}, that join GPUs and scripting languages
in one programming environment.

Before we move on, however, let us comment on a practicality: In
today's GPU programming environments (OpenCL, CUDA), all the host
computer is required to do is submit work to the compute device at a
certain rate, typically around 1000 Hz. As long as the scripting-based
host program can maintain this rate, there is no loss in performance.

PyOpenCL and PyCUDA can be used in a large number of roles, for example as a
prototyping and exploration tool, to help with optimization, as a
bridge to the GPU for existing legacy codes (in Fortran, C, or other
languages), or, perhaps most excitingly, to support an unconventional
\emph{hybrid way of writing high-performance codes}, in which a
high-level controller generates and supervises the execution of
low-level (but high-performance) computation tasks to be carried out
on varied GPU- or GPU-based computational infrastructure.

Scripting languages already excel at text processing and are routinely
used for this task at extreme scales, as exemplified by their use in
the generation of HTML pages. This already makes them a good choice
for the textual part of code generation. A number of further points
contribute to making the programming environment created by joining
GPUs and scripting greater than just the sum of its two parts. First,
scripting languages lend themselves to very clean programming
interfaces, with seamless (but invisible) error reporting and
automatic resource management. In addition, scripting languages are
very suited to creating abstractions, and Python especially follows a
``batteries included'' approach that puts many of these abstractions
directly within a user's reach. PyOpenCL and PyCUDA strive to make
ideal use of these characteristics. They are fully documented, and
also come with ``batteries included''--for instance, users do not have to reinvent
vectors, arrays, reductions or prefix sums. Both packages also cache
compiler output, to support RTCG and retain the development ``feel''
of a scripting language.

The Python programming language \citep{vanrossum_python_1994}
is well-suited for such packages for a number of reasons:
\begin{itemize}
  \item The existence of a mature array abstraction (\texttt{numpy}
  \citep{oliphant_numpy_2006}) facilitating (in-process) transport and
  manipulation of bulk numerical data.
  \item The large ecosystem of software that has sprung up around
  \texttt{numpy}.
  \item Its main-stream syntax and language-features, which make
  the language easy to learn while not impeding more advanced use.
\end{itemize}
Other languages may certainly be just as suitable.

In concluding this argument for GPUs, scripting and RTCG, let us
remark that the packages introduced here are distributed under the
liberal MIT license and are available at the URLs
\url{http://mathema.tician.de/software/pyopencl} (or
\texttt{/pycuda}). A mailing list, a wiki, and a number of contributed
computational add-on packages are available. Both packages are
routinely used on Windows, OS X, and Linux.

\section{Results: RTCG for Discontinuous Galerkin}
\label{sec:results}
\begin{figure}
  a)
  \[
    \hat n \cdot( F-F^{\ast})_E:=
    \frac 12
    \left[
      \hat n \times (\jump{H} - \alpha \hat n \times \jump{E})
    \right]
  \]
  b)
\begin{lstlisting}
flux = 1/2*cross(normal, h.int-h.ext
      -alpha*cross(normal, e.int-e.ext))
\end{lstlisting}
  c)
\begin{lstlisting}
a_flux += (
    (((val_a_field5 - val_b_field5)*fpair->normal[2] 
      - (val_a_field4 - val_b_field4)*fpair->normal[0])
     + val_a_field0 - val_b_field0)*fpair->normal[0] 
    - (((val_a_field4 - val_b_field4) *fpair->normal[1] 
        - (val_a_field1 - val_b_field1)*fpair->normal[2])
      + val_a_field3 - val_b_field3) * fpair->normal[1]
    )*value_type(0.5);
\end{lstlisting}
  \caption{
    Three representations of a (partial) numerical flux for the
    Maxwell equations. a) shows the mathematical specification as
    first given in \citep{mohammadian_computation_1991}. b) shows the
    Python code used to instruct the solver. c) shows a fraction
    (about one sixth) of the C code ultimately generated by the solver
    to implement the flux in a).
  }
  \label{fig:flux-code}
\end{figure}

Having introduced run-time code generation as a way of addressing the
challenges encountered in Section \ref{sec:gpu-dg}, we will refocus on
some of the specific benefits that RTCG brings in the context of a
discontinuous Galerkin solver and discuss a few of the achieved
results. For definiteness, we will be discussing results obtained 
using the solver ``hedge'', which was built to explore and develop the
ideas in this article.

The first example emphasizes the impact of RTCG on the ability to
write maintainable software with reasonable user interfaces. In
particular, we will demonstrate the user interface that our DG solver
code uses to specify numerical flux terms--the terms $(n \cdot F^*)$
in \eqref{eq:strong-dg}. Figure \ref{fig:flux-code} shows three
representations of a (partial) numerical flux for the Maxwell
equation. First, Figure \ref{fig:flux-code}a) shows the mathematical
notation as one might find in a scientific article. Next, Figure
\ref{fig:flux-code}b) shows the Python code that a user might need to
write to capture the numerical flux expression of a) in our solver.
The final part c) of the figure shows a fraction of the generated
code. What this seeks to demonstrate is that a high-performance,
low-level, scalar C-language representation can easily be generated
from a high-level, vectorial statement in a scripting language. It is
obvious that the code in Figure \ref{fig:flux-code}b) is much easier to
check for correctness than the resulting C code. Nonetheless, even
textbooks such as \citep{hesthaven_nodal_2007} contain code like 
that of \ref{fig:flux-code}c) for
demonstration purposes. By using RTCG, in many situations it becomes a
rather easy proposition to enable the user to write maintainable,
transparent code, and still obtain all the performance of a
program that would have previously required a rather large amount of
manual labor and checking.

\begin{figure}
  \sidecaption
  \includegraphics[width=0.6\textwidth]{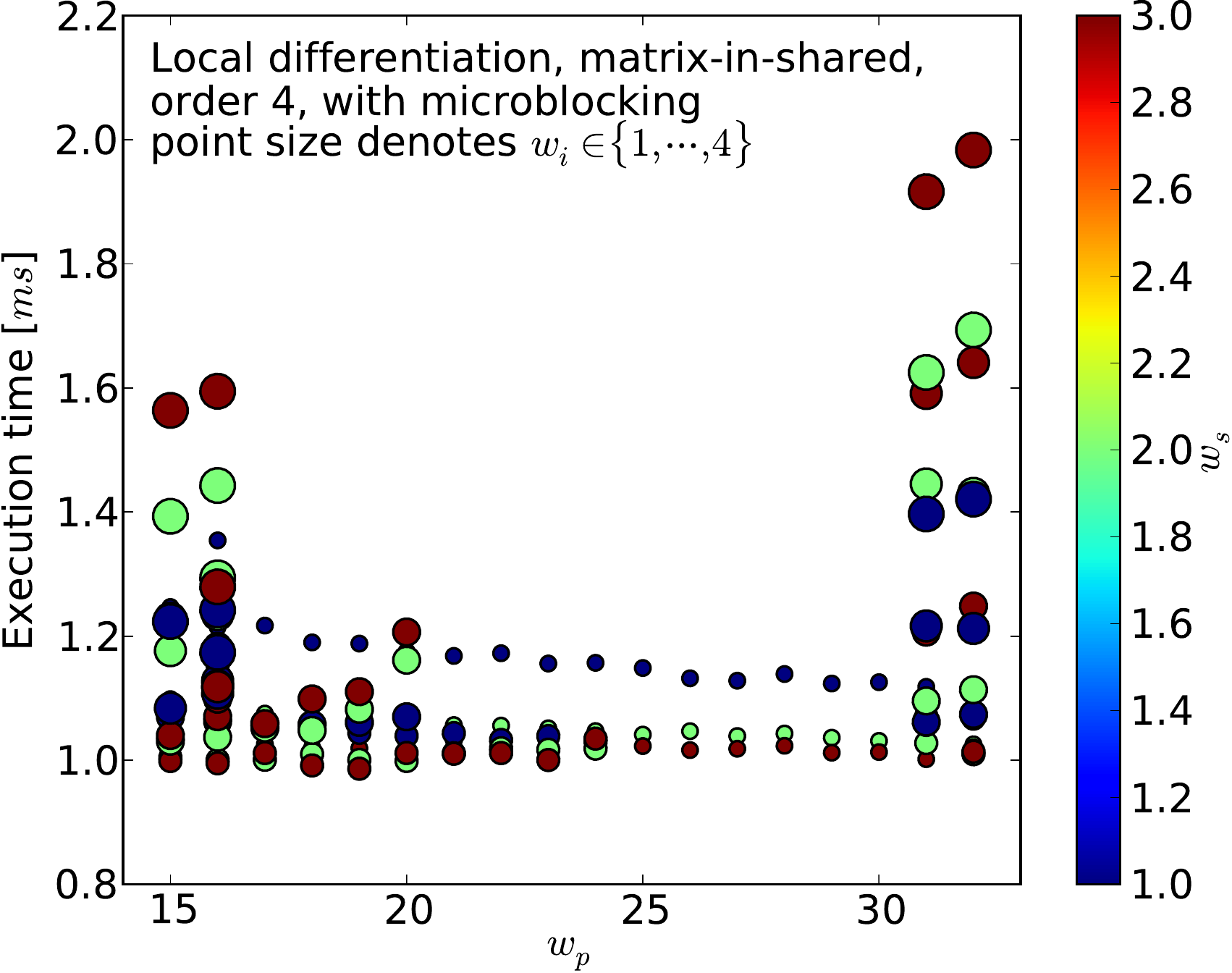}
  \caption{
    Sample tuning study for local differentiation
    on fourth-order elements with microblocking enabled, showing time
    spent for a constant amount of work depending on the values $w_s$,
    $w_p$ and $w_i$ introduced in Section \ref{sec:gpu-dg}.
  }
  \label{fig:tuning-study}
\end{figure}

Further, our solver obviously makes extensive use of automated tuning.
Figure \ref{fig:tuning-study} shows results from a particular tuning
run attempting to optimize the parameters $w_s$, $w_p$ and $w_i$
introduced in Section \ref{sec:gpu-dg} for element-local
differentiation. The vertical axis of the plot shows timing
information, and each of the dots in the plot represents a particular
timing run. The same amount of numerical work was done for each of the
dots, yet surprisingly, the final performance varied by more than a
factor of 2 depending on parameter choice. In addition, there is
little observable regularity in the graph, which seems to limit the
amount of success that any given heuristic might have in predicting
this behavior. There is almost no other option besides automated
tuning to find an at least somewhat optimal combination within 
this parameter space. Also note that any performance gain in this part
of the operator has a rather large impact on the performance of the
method as a whole--element-local differentiation is the asymptotically
most work-intensive part of a DG operator.

In addition to this application of automated tuning in the
determination of a parallel work decomposition, our solver also
applies this technique in finding memory layouts and flux gather
granularities. Further, by virtue of code generation, it naturally
benefits from being able to ``hard-code'' certain variable values such
as matrix sizes, polynomial degrees, or loop trip counts.

In the following, we will present a number of overall performance
results for our solver on an Nvidia GTX 280, to confirm that a
high-performance solver can be written using the techniques described.
Unless otherwise specified, all performance numbers are based on the
wall clock time from the beginning of one time step to the beginning
of the next, including RK4 timestepping.  Timings were averaged over a
run of 100 (CPU) or several hundred (GPU) time steps to minimize the
influence of timing transients. Timings were observed to be consistent
across runs, even when using automated tuning.

\begin{figure}
  \sidecaption
  \includegraphics[width=0.6\textwidth]{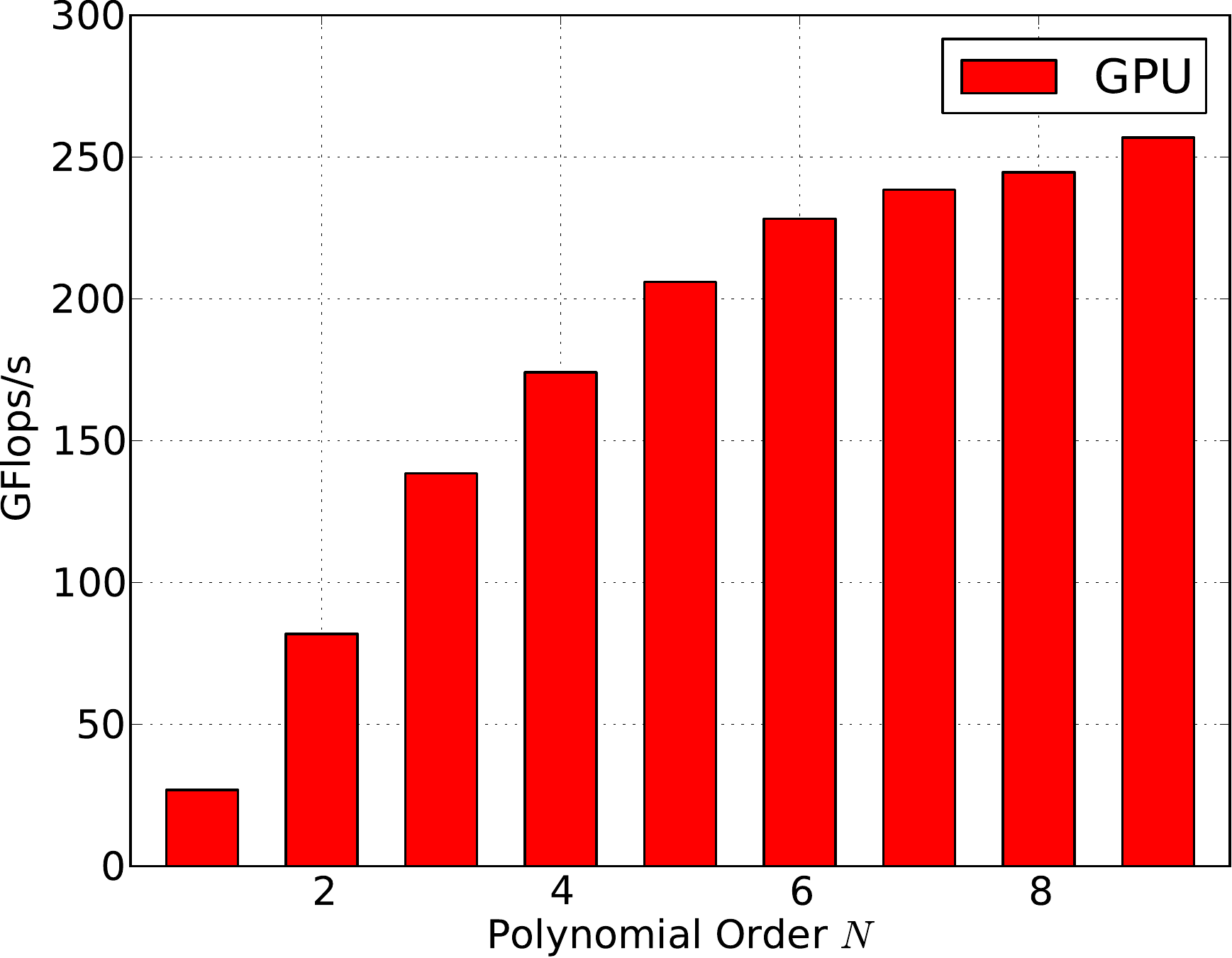}
  \caption{
    Floating point performance in GFlops/s achieved by our auto-tuning
    solver on a large 3D Maxwell problem in single precision
    on an Nvidia GTX 280.
    Performance is calculated by measuring wall time from one time
    step to the next and dividing the number of flops performed
    (including timestepping) by this value.
  }
  \label{fig:gflops-order}
\end{figure}

Figure \ref{fig:gflops-order} shows overall performance expressed in
billions of floating point operations per second (GFlops/s), measured
by counting flops over a time step and dividing by the duration of
that same time step. This is a reasonable (and reproducible)
measurement, because unlike for many other numerical methods, the
number of flops required for simplicial DG is relatively uniquely
determined. Note that because the measurement corresponds to an
average, individual components of the method (such as element-local
differentiation/lift) achieve significantly higher flop rates.  Since
the elementwise dense linear operators asymptotically (as
$N\to\infty$) determine the run time, it may be reasonable to relate
the measured performance to that achieved by dense matrix-matrix
multiplies on this architecture. The best results achieved on an SGEMM
workload on large, square matrices hover around 350 GFlops/s. It is
therefore remarkable that our method achieves 250 GFlops/s on much
less benignly shaped matrices, also taking into account that the
method does much more varied work than simple matrix-matrix multiplies.

We would also like to comment on the progression of performance
results as we vary $N$ in Figure \ref{fig:gflops-order}, and in
particular the rapid rise in performance from $N=1$ to $N=3$. We
mentioned earlier that optimal results for $N=3,\dots,5$ were an explicit goal of this
work. Many of the finer tuning points of past sections (such as
microblocking and work item coarsening) become rather unnecessary at
$N\ge 6$ (because matrix sizes have grown significantly, and therefore
enough work is available within each element). Compared to a simpler
code (such as the one described in \citep{kloeckner_solving_2011}),
it is precisely these optimizations that lead to large gains at
$N=3,\dots, 5$.

\begin{figure}
  \sidecaption
  \includegraphics[width=0.6\textwidth]{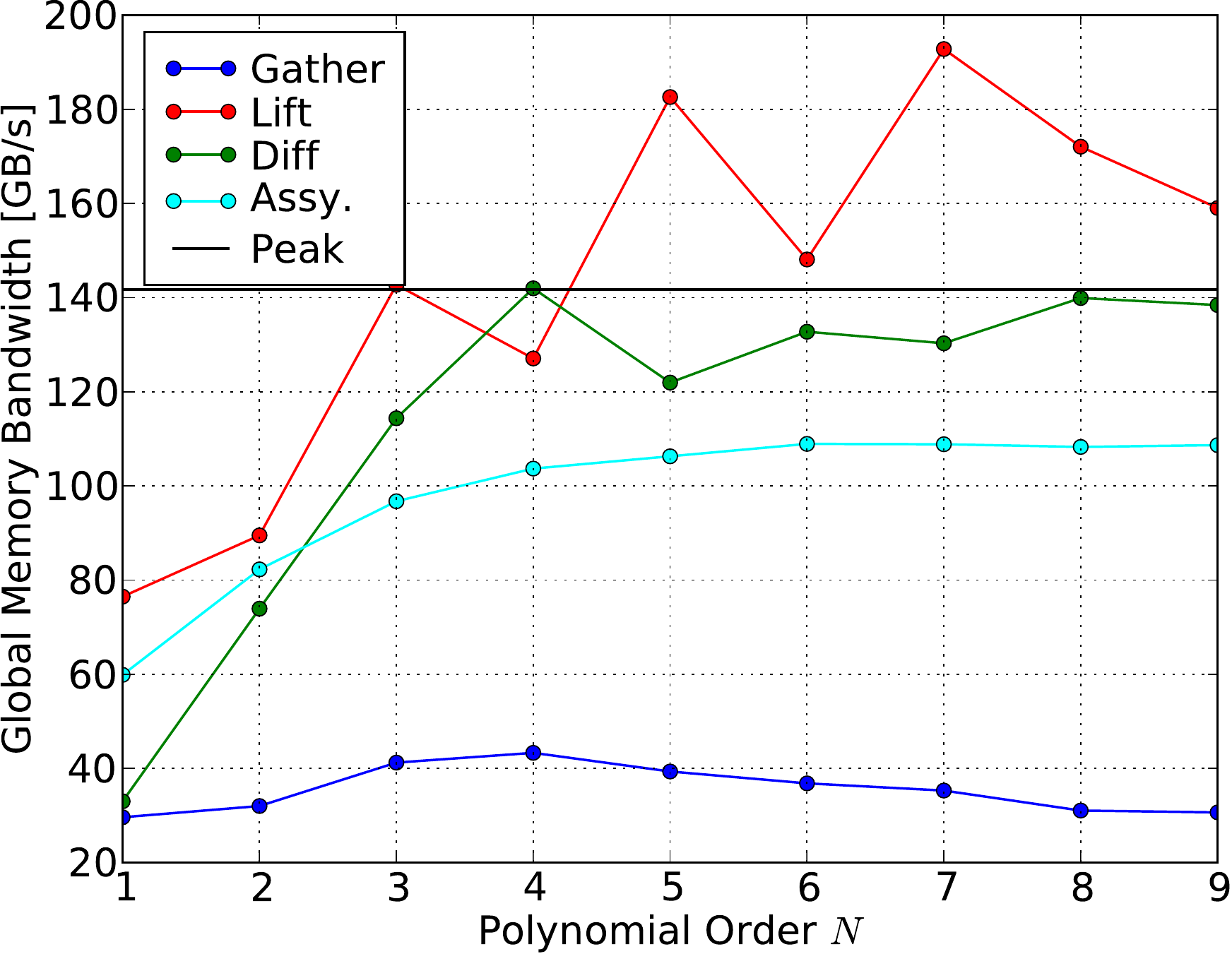}
  \caption{
    Memory bandwidths in GB/s achieved by each part of the DG operator
    on an Nvidia GTX 280.  The peak memory bandwidth published by the
    manufacturer is 141.7 GB/s.  Values exceeding peak bandwidth are
    believed to be due to the presence of a texture cache.
  }
  \label{fig:mem-bandwidth}
\end{figure}

It is interesting to correlate the achieved floating point bandwidth
from Figure \ref{fig:gflops-order} with the bandwidth reached for
transfers between the processing cores and global memory, shown in
Figure \ref{fig:mem-bandwidth}. We have obtained these numbers by
counting the number of bytes fetched from global memory either
directly or through a texture unit in each component of the method.
The published theoretical peak memory bandwidth of the card on which
this experiment was performed is 141.7 GB/s, shown as a black
horizontal line. Perhaps the most striking feature here is that the
calculated memory bandwidth sometimes transcends this theoretical
peak. We attribute this phenomenon to the presence of various levels
of texture cache.  Its occurrence is especially pronounced in the case
of flux lifting, and it should perhaps be sobering that the other
parts of the DG operator do not manage the same feat.  In any case,
flux lifting uses the fields-on-chip strategy, and therefore fetches
and re-fetches the rather small matrix $L$, making large amounts of
data reuse a plausible proposition.  Aside from this surprising
behavior of flux lifting, it is both interesting and encouraging to
see how close to peak the memory bandwidth for element-local
differentiation gets. As a converse to the above, this makes it likely
that the operation does not get much use out of the texture cache in
most situations. It does imply, however, that rather impressive work
was done by Nvidia's hardware designers: The theoretical peak global
memory bandwidth can very nearly be attained in real-world
computations.  Next, the fact that the flux-gather part of the
operator achieves rather low memory throughput is not too
surprising--the access pattern is (and, for a general grid, has to be)
rather scattered, decreasing the achievable bandwidth. Lastly,
operator assembly, which computes linear combination of vectors,
consists mainly of global memory fetches and stores. It seems likely
that ancillary operations such as index calculations, loop overhead
and bounds checks drive this component's shortfall from peak memory
bandwidth.

It is worth noting that one would not initially expect a matrix-matrix
workload like DG to be memory-bound, at least at high polynomial
degrees $N$. After all, such workloads do offer large amounts of
arithmetic intensity to keep floating point units busy. On the other
hand, it is worth keeping in mind that there is simply \emph{so much}
floating point power available on GPU-like chips that it is quite
unlikely that a code like DG might get to the point of actually being
limited by it. As such, it is reasonable, in our view, to expect that
for the foreseeable future, the limiting factor for most DG-like
algorithms will in fact remain memory bandwidth, as evidenced by
Figure \ref{fig:mem-bandwidth}.

Another issue that frequently draws questions is that of the support
of double precision within GPU-like devices. Marketing pressure in
this area has led GPU manufactures to increase the ratios of the
number of double precision (DP) units to the number of single
precision (SP) units. Current high-end offerings hover between a
factor of 1/2 and 1/4, where this feature is often used to
differentiate between `consumer-grade' and `professional-grade'
hardware. We would like to remark that in bandwidth-bound
applications, there is no reason to expect a DP code to go any faster
than half as fast as an equivalent SP code, for the simple reason that
DP numbers are exactly twice as big as SP numbers, and therefore
require twice as much memory bandwidth. In addition, DP requires twice
as much on-chip memory to obtain equivalent levels of data reuse--an
amount that simply might not be available. With respect to DG, we
observe that at low $N$ (e.g. $N=1,2$), the ratio
(DP~GFlops/s)/(SP~GFlops/s) is about a factor 1/2, as the algorithm is
completely bandwidth bound. As $N$ increases, it approaches the
above-mentioned ratio of (available~DP~units)/(available~SP~units),
which further substantiates the conjecture made above that the code is
``underway'' to being compute-bound.

\begin{figure}
  \sidecaption
  \includegraphics[width=0.6\textwidth]{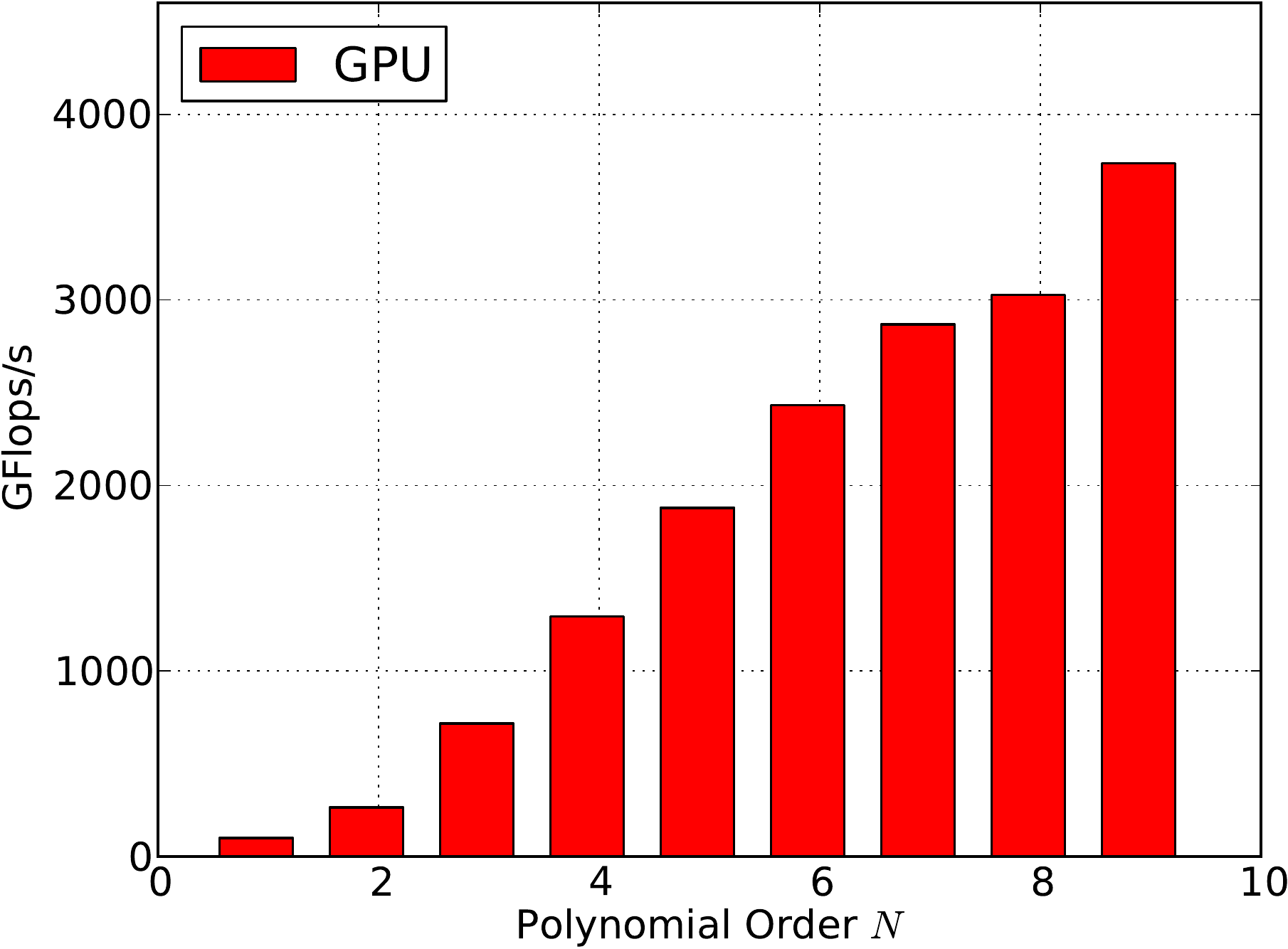}
  \caption{
    Floating point performance in GFlops/s achieved by our auto-tuning
    solver on a very large 3D Maxwell problem on 16 Nvidia T10 GPUs
    (parts of an Nvidia S1070 compute server) in single precision.
    Performance is calculated by measuring wall time from one time
    step to the next and dividing the number of flops performed
    (including timestepping) by this value.
  }
  \label{fig:mpi-perf}
\end{figure}

Lastly, we would like to comment on Figure \ref{fig:mpi-perf}, which
illustrates the performance of our solver in GFlops/s at various
polynomial orders $N$ on a cluster of 16 Nvidia Tesla T10 GPUs. Two
features of this plot are immediately noteworthy. First, computational
performance approaches 25\% of the overall machine peak at $N=9$ with
nearly four teraflops/s. It is remarkable that such performance is
achievable on a cluster that costs a small fraction of the large
machines whose hallmark such performance was previously. Second, it is
also obvious where the distributed-memory inter-node communication
(via MPI in this case) is taking its toll, as one may, in principle,
directly compare the shape of Figure \ref{fig:gflops-order} with that
of Figure \ref{fig:mpi-perf}. It is obvious that there is a much steeper
performance dropoff in the parallel run as $N$ decreases than there is
in the sequential performance data. This is owed to the fact that high
orders are significantly heavier on element-local volume work (which
scales as $N^3$), than on communication-heavy work dominated by
degrees of freedom on faces (which scales as $N^2$). Thus, as there is
more communication work compared to local compute work, the method
incurs larger communication overhead. This is (in our opinion) quite
expected, and it should be noted that even at $N=5$, our code nearly
achieves a still very respectable 2 teraflops/s on this cluster. This
also contains an important message about the parallelization of DG,
which holds true at both the distributed-memory and the shared-memory
scale: High
polynomial orders $N$, along with all their other benefits, also much
improve the parallelizability of the method.

\section{Conclusions}
\label{sec:conclusions}

In this article, we have shown that high-order DG methods can reach
double-digit percentages of published theoretical peak performance
values for the hardware under consideration. This speed
increase translates directly into an increase of the size of the
problem that can be treated using these methods.  A single compute
device can now do work that previously required a roomful of computing
hardware. Alternatively, a cluster of machines equipped with these
cards can run simulations that were previously outside the reach of
all but the largest supercomputers.  This lets the size and complexity
of simulations that researchers can afford on a given hardware budget
jump significantly.

We find that GPU-DG is far more economical to run at medium to large
scales than CPU-DG. In our opinion, this is due to the fact that the
computational structure of the method, with its two levels of ``element''
and ``individual degree of freedom'', is very well-suited to the GPU a
priori--better even than finite-difference methods, which are often
cited as a ``GPU poster child''. Through the use of the auto-tuning technology
described in this article along with a number of further tricks
discussed in detail in \citep{kloeckner_nodal_2009}, we have shown
that rather good performance and machine utilization can be achieved
by GPUs in DG-like workloads.

In addition to highlighting our work on GPU-DG, this article also
serves to introduce the reader to the idea that scripting languages
and GPUs make a good team. Beyond the core benefit of enabling
run-time code generation, they also facilitate a clear separation of
the code into `administrative' and `computational' parts. Such a
separation contributes to code clarity and helps make code more
maintainable.

As we continue to explore the benefits of GPUs for DG and DG-like
workloads, we will be focusing on areas such as adaptivity in both space
and time, nonlinear equations, and the use of curvilinear geometries, as
well as much larger scaling of GPU-DG. Initial work on these matters can
be found in the articles
\citep{kloeckner_viscous_2011,burstedde_extreme-scale_2010,warburton_low-storage_2010}.

We believe that GPU-DG will have a bright future, with many more
applications benefiting from the ease with which large-scale
time-domain simulations can be be performed using DG, and we hope that
our work has helped and will help application scientists use DG computations
in their role as part of the `third pillar of science'.

\subsection*{Acknowledgments}

AK's research was partially funded by AFOSR under contract number
FA9550-07-1-0422, through the AFOSR/NSSEFF Program Award
FA9550-10-1-0180 and also under contract DEFG0288ER25053 by the
Department of Energy. TW acknowledges the support of AFOSR under
grant number FA9550-05-1-0473 and of the National Science Foundation
under grant number DMS 0810187. JSH was partially supported by AFOSR,
NSF, and DOE.  The opinions expressed are the views of the authors.
They do not necessarily reflect the official position of the funding
agencies.

\bibliographystyle{abbrvnat}
\bibliography{article}

\end{document}